\documentstyle[12pt]{article}

\newcommand{\be}{\begin{equation}}
\newcommand{\ee}{\end{equation}}
 
\newcommand{\bea}{\begin{eqnarray}}
\newcommand{\eea}{\end{eqnarray}}

\newcommand{\NP}[1]{Nucl. Phys.\ {\bf #1}\ }
\newcommand{\PL}[1]{Phys. Lett.\ {\bf #1}\ }
\newcommand{\NC}[1]{Nuovo Cimento\ {\bf #1}\ }

\newcommand{\PR}[1]{Phys. Rev.\ {\bf #1}\ }
\newcommand{\PRL}[1]{Phys. Rev. Lett.\ {\bf #1}\ }

\newcommand{\IJMP}[1]{Int. J. Mod. Phys.\ {\bf #1}\ }

\newcommand{\CQG}[1]{Class. Quantum Grav.\ {\bf #1}\ }

\newcommand{\JMP}[1]{J. Math. Phys.\ {\bf #1}\ }
 
\def\lsim{\raise0.3ex\hbox{$<$\kern-0.75em\raise-1.1ex\hbox{$\sim$}}}
\def\gsim{\raise0.3ex\hbox{$>$\kern-0.75em\raise-1.1ex\hbox{$\sim$}}}

\newcommand{\CR}{\nonumber \\}

\newcommand{\lm}{\lambda}

\newcommand{\ie}{{\it i.e. }}
\newcommand{\eg}{{\it e.g. }}

\begin{document}
\begin{titlepage}
\begin{flushright}
{\large \bf HU-SEFT R 1995-18}
\end{flushright}
\vskip 2cm
\begin{center}

{\Large \bf
Spinning Relativistic Particle in an External Electromagnetic Field}
\vskip .4in
 
M. Chaichian$^{a,b}$, R. Gonzalez Felipe$^{b}$ and 
D. Louis Martinez$^{c}$
\\[.15in]
 
{\em $^{a}$Laboratory for High Energy Physics, Department of Physics}

{\em $^{b}$Research Institute for High Energy Physics}\\[.15in]

{\em 
P.O. Box 9 (Siltavuorenpenger 20 C)\\
FIN-00014 University of Helsinki, Finland}\\[.15in]

{\em $^{c}$Department of Physics, University of Manitoba\\[.15in]
 Winnipeg MB R3T 2N2, Canada }

\end{center}
 
\vskip 3cm
 
\begin{abstract}
The Hamiltonian formulation of the motion of a spinning 
relativistic particle in an external electromagnetic field is considered.
The approach is based on the introduction of new coordinates and their
conjugated momenta to describe the spin degrees of freedom together with 
an appropriate set of constraints in the Dirac formulation. For particles
with gyromagnetic ratio $g=2$, the equations of motion do not predict
any deviation from the standard Lorentz force, while for $g \neq 2$
an additional force, which corresponds to the magnetic dipole force,
 is obtained.   
\end{abstract}
\end{titlepage}

\newpage
\renewcommand{\thepage}{\arabic{page}}
\setcounter{page}{1}
\setcounter{footnote}{1}
 

The description of the motion of a spinning relativistic particle in 3+1
dimensions has been a subject of research for many years. Equations 
describing the evolution of the classical spin in a uniform and static
external electromagnetic field have been proposed by several authors
\cite{frenkel,BMT}. Modifications of the Lorentz force law for spinning
particles moving in a slowly varying external electromagnetic field 
together with some generalizations for the equations describing the spin
precession in this case have also been suggested 
\cite{aharonov}-\cite{holten}.
Some Lagrangian and Hamiltonian formulations, which allow one to obtain 
the equations of motion from an action principle, have been proposed
\cite{barducci}-\cite{costella}. Most of them rely on the use of Grassmann
variables \cite{barducci}-\cite{rietdijk}. In \cite{khriplovich} the 
analogy with the relativistic quantum equations was used as a guiding 
principle. Recently, in Ref. \cite{bander} a canonical procedure  has been 
proposed based on a Routhian 
function \cite{landau} from which the equations of motion for a spinning 
relativistic particle coupled to an external electromagnetic and a
gravitational field can be derived. Also, in \cite{costella} a Lagrangian
formulation based on classical commuting variables was proposed.
 It is however amusing that there is, even today, an ongoing 
debate as to the torque and force on such particles.

In this letter we shall consider the Hamiltonian formulation for the
description of the motion of a spinning relativistic particle in 3+1 
space-time dimensions in an homogeneous constant electromagnetic field.
The model is constructed by analogy with the one presented in
\cite{chaichian} for anyons in 2+1 dimensions. It uses the idea of 
expressing the antisymmetric spin tensor in terms of some new coordinates
and their conjugated momenta as in the case of the orbital angular 
momentum tensor. We start with a Hamiltonian formulation of the problem
by introducing the corresponding constraints in the theory. The Lagrangian
can then be derived by using the usual Legendre transformation. We hope 
that the present model will provide an alternative formulation to the 
ones discussed in the literature and mentioned above.

Let the position of the relativistic particle be described by the four 
vector $x^\mu$ ($\mu=0,1,2,3$), and the spin of the particle by the 
antisymmetric spin matrix $S_{\mu\nu}$. We shall introduce the auxiliary
coordinates $n^\mu$ to describe the spin. Therefore, in the phase space
we work with the configuration space variables $x^\mu\ ,\ n^\mu$ and their 
conjugated momenta $p_\mu\ ,\ p^{(n)}_\mu$. We can write the following 
Poisson bracket (PB) relations:
\bea
\{x_\mu\ ,\ p_\nu\}=-g_{\mu\nu}\ ,\ \{x_\mu\ ,\ x_\nu\}=
\{p_\mu\ ,\ p_\nu\}=0\ , \CR
\{n_\mu\ ,\ p^{(n)}_\nu\}=-g_{\mu\nu}\ ,\ \{n_\mu\ ,\ n_\nu\}=
\{p^{(n)}_\mu\ ,\ p^{(n)}_\nu\}=0\ ,
\label{1}
\eea   
where the metric tensor $g_{\mu\nu}=diag(1,-1,-1,-1)$.

Let us define the antisymmetric spin matrix $S_{\mu\nu}$ as follows:
\be
S_{\mu\nu}=n_\mu p^{(n)}_\nu-n_\nu p^{(n)}_\mu\ .
\label{2}
\ee
The total angular momentum matrix $M_{\mu\nu}$ can thus be written as
\be
M_{\mu\nu}=L_{\mu\nu} + S_{\mu\nu}\ ,
\label{3}
\ee
where $L_{\mu\nu}=x_\mu p_\nu - x_\nu p_\mu$\ .

It is straightforward to verify that the four-vector $p_\mu$ and the tensor
$M_{\mu\nu}$ satisfy the Poincar\'e algebra
\bea
\{p_\mu\ ,\ p_\nu\}=0\ ,\CR
\{M_{\mu\nu}\ ,\ p_\lm\}=-g_{\mu\lm} p_\nu+g_{\nu\lm} p_\mu\ ,\CR
\{M_{\mu\nu}\ ,\ M_{\lm\rho}\}=-g_{\mu\lm} M_{\nu\rho}-g_{\nu\rho} M_{\mu\lm}
+g_{\mu\rho} M_{\nu\lm}+g_{\nu\lm} M_{\mu\rho}\ . 
\label{4}
\eea

Now we would like to consider the interaction of a spinning particle
with gyromagnetic ratio $g=2$, electric charge $e$ and mass $m$ 
with a uniform static external electromagnetic field. A vector potential
describing such a field can be expressed as 
\be
A_\mu=-\frac{1}{2}F_{\mu\nu} x^\nu \ ,
\label{5}
\ee
where $F_{\mu\nu}$ is the electromagnetic field tensor. 

In order to describe the motion of such a particle we shall follow 
a Hamiltonian
formulation analogous to the one presented in \cite{chaichian} and based
on the imposed constraints and the algebraic properties of the system.

Let us consider the following extended (with respect to the Poincar\'e)
 algebra:

\bea
\{\bar{\pi}_\mu\ ,\ \bar{\pi}_\nu\}=e F_{\mu\nu}\ ,\CR
\{M_{\mu\nu}\ ,\ \bar{\pi}_\lm\}=-g_{\mu\lm} \bar{\pi}_\nu+
g_{\nu\lm} \bar{\pi}_\mu\ ,\CR
\{M_{\mu\nu}\ ,\ M_{\lm\rho}\}=-g_{\mu\lm} M_{\nu\rho}-g_{\nu\rho} M_{\mu\lm}
+g_{\mu\rho} M_{\nu\lm}+g_{\nu\lm} M_{\mu\rho}\ ,\CR
\{F_{\mu\nu}\ ,\ \bar{\pi}_\lm\}=0\ ,\CR
\{F_{\mu\nu}\ ,\ F_{\lm\rho}\}=0\ ,\CR
\{M_{\mu\nu}\ ,\ F_{\lm\rho}\}=-g_{\mu\lm} F_{\nu\rho}-g_{\nu\rho} F_{\mu\lm}
+g_{\mu\rho} F_{\nu\lm}+g_{\nu\lm} F_{\mu\rho}\ ,
\label{6}
\eea
where $\bar{\pi}_\mu=p_\mu+e A_\mu$. 

It is not difficult to prove that the operator
\be
C=\bar{\pi}^2- e F_{\mu\nu} M^{\mu\nu}
\label{7}
\ee
is the Casimir operator of the algebra (\ref{6}). By using the relation
$$
F_{\mu\nu} L^{\mu\nu} = 4 (p A)\ ,
$$
Eq.(\ref{7}) can be written in the form:
\be
C=\pi^2- e F_{\mu\nu} S^{\mu\nu}\ ,
\label{8}
\ee
with $\pi_\mu=p_\mu-e A_\mu$. By fixing the value of $C$ to be equal to
$m^2$ we can then take the quantity
\be
\Phi=\pi^2- e F_{\mu\nu} S^{\mu\nu}-m^2
\label{9}
\ee
as a first-class constraint \cite{dirac} in the Hamiltonian formulation.

Next we should require that in the rest frame of the particle the spin
tensor $S_{\mu\nu}$ has only three independent components.
 This is usually done by imposing the conditions:
\be
S_{\mu\nu} \dot{x}^\nu =0\ ,
\label{10}
\ee
where $\dot{x}^\mu$ is the four-velocity and the dot denotes the 
derivative with respect to the proper time $\tau$. 
Since we are considering the
Hamiltonian formalism we should employ the momenta instead
of the velocities. In an external electromagnetic field we
thus expect the use of $\pi^\mu$ instead of $\dot{x}^\mu$. On the other
hand, it is easy to see that from the definition of the spin tensor 
(\ref{2}) it is sufficient to impose the constraints
$$
(\pi p^{(n)})=0\ ,\ (\pi n)=0\ ,
$$
in order to guarantee that

\be
S_{\mu\nu} \pi^\nu =0\ .
\label{11}
\ee

Therefore, in our Hamiltonian formulation we shall consider the 
following set of primary constraints:
\bea
\Phi=\pi^2- e F_{\mu\nu} S^{\mu\nu}-m^2\ ,\CR
\varphi_1=(\pi  p^{(n)})\ , \CR
\varphi_2=(\pi  n)\ .
\label{12}
\eea
   
Since
\bea
\{\Phi , \varphi_1\}=0\ ,\CR   
\{\Phi , \varphi_2\}=0\ ,\CR
\{\varphi_1 , \varphi_2\}=\pi^2+\frac{e}{2} F_{\mu\nu} S^{\mu\nu} ,
\label{13}
\eea
the constraint $\Phi $ is a first-class constraint, while $\varphi_1$
and $\varphi_2$ form a pair of second-class constraints. We shall define 
the total Hamiltonian of the system as
\be
H= \Lambda \Phi +\lm_1\varphi_1+\lm_2\varphi_2\ ,
\label{14}
\ee
where $\Lambda, \lm_1$ and $\lm_2$ are Lagrange multipliers and
the canonical Hamiltonian $H_{can}$ is taken to be equal to zero.

From the consistency conditions on the primary constraints (\ref{12}), 
\ie from the conditions
\bea
 \dot{\Phi}= \{\Phi , H\}=0\ ,\CR   
 \dot{\varphi_1}=\{\varphi_1 , H\}=\lm_2 (\pi^2+\frac{e}{2} F_{\mu\nu} 
S^{\mu\nu})=0 \ ,\CR   
 \dot{\varphi_2}=\{\varphi_2 , H\}=-\lm_1 (\pi^2+\frac{e}{2} F_{\mu\nu} 
S^{\mu\nu})=0\ ,   
\label{15}
\eea
it follows that 
\be
\lm_1=\lm_2=0
\label{16}
\ee
on the constrained surface and that no new constraints appear in the Dirac 
algorithm. The Lagrange multiplier $\Lambda$  remains undetermined.

Notice that the number of physical degrees of freedom in configuration 
space is equal to $8-1-2/2=6$. It is thus expected that 3 of these 
correspond to the position of the relativistic particle and the other
3 to the components of the spin. 

It is useful to define the spin four-vector through the relation:
\\
\be
S_\mu=\frac{1}{2}\varepsilon_{\mu \alpha \beta \gamma }
\frac{\pi^\alpha}{\sqrt{\pi^2}} S^{\beta \gamma }\ ,
\label{17}
\ee
\\
which in the absence of an external electromagnetic field reduces to the
usual Pauli-Lubanski four-vector \cite{itzykson}. 
Using Eqs. (\ref{11}) and (\ref{17}) it is not difficult to prove that 
the spin 
matrix $S_{\mu\nu}$ can be written on the constrained surface as follows:
\\
\be
S_{\mu\nu}=\varepsilon_{\mu \nu \alpha \beta} S^{\alpha}
\frac{\pi^\beta}{\sqrt{\pi^2}} \ .
\label{18}
\ee
\\
It also follows that
\\
\be
S^2\equiv S_\mu S^\mu=-\frac{1}{2} S_{\mu\nu}S^{\mu\nu}\ .
\label{19}
\ee
\\

The Dirac-Hamilton equations of motion which follow from the PB (\ref{1})
and the Hamiltonian (\ref{14}) read
\bea
\dot{x}^\mu=2\Lambda \pi^\mu\ ,\CR
\dot{n}^\mu=2\Lambda e F^{\mu\nu} n_\nu\ ,\CR
\dot{p}_\mu=\Lambda e F_{\mu\nu} \pi^\nu\ ,\CR
\dot{p}^{(n)}_\mu=2\Lambda e F_{\mu\nu} p^{(n) \nu}\ ,
\label{20}
\eea
and they should hold together with the constraints (\ref{12}). 

Let us now rewrite these equations in a more familiar way. First notice
that
$$
\dot{\pi}_\mu=\dot{p}_\mu+\frac{e}{2}F_{\mu\nu}\dot{x}^\nu\ .
$$
Using Eqs.(\ref{20}) and choosing $\Lambda=1/2m$ we obtain
\be
\ddot{x}_\mu=\frac{e}{m}F_{\mu\nu} \dot{x}^\nu \ .
\label{21}
\ee

Also from the definition of the spin matrix (cf. Eq.(\ref{2})) and 
the equations of motion (\ref{20}) it follows that
\\
\be
\dot{S}_{\mu\nu}=-\frac{e}{m} F^{\alpha \beta }\left( S_{\mu\alpha}
g_{\nu \beta }+S_{\nu \beta } g_{\mu \alpha }\right) \ .
\label{22}
\ee
\\

Eqs.(\ref{21}) and (\ref{22}) imply that the Frenkel conditions (\ref{10}) 
are preserved in proper time, \ie 
\be
\frac{d}{d\tau}\left( S_{\mu\nu} \dot{x}^\nu\right) =0\ .
\label{23}
\ee
Also from (\ref{19}) and (\ref{22}) it follows that the value of the spin 
is a  constant of motion:
\be
\frac{d}{d\tau}\left( S^2\right) =0\ .
\label{24}
\ee

Finally, Eq.(\ref{22}) can be rewritten in terms of the spin vector 
defined in (\ref{17}). We have
\\
\be
\dot{S}_{\mu}=\frac{e}{m} F_{\mu \nu } S^{\nu}\ ,
\label{25}
\ee
\\
which coincides with the Bargmann-Michel-Telegdi equation \cite{BMT} for
spinning particles with $g=2$.

For completeness, let us derive the Lagrange function corresponding
to the total Hamiltonian (\ref{14}). The derivation is very similar to 
the one presented in \cite{chaichian}. The first step is to express the 
Lagrange multipliers $\Lambda$, $\lm_1$ and $\lm_2$ in terms of
the generalized coordinates and velocities. This can be achieved by 
using the relations
\bea
\dot{x}_\mu=2\Lambda \pi_\mu\ + \lm_1 p^{(n)}_\mu+\lm_2 n_\mu,\CR
\dot{n}_\mu=2\Lambda e F_{\mu\nu} n^\nu\ +\lm_1 \pi_\mu\ , 
\label{26}
\eea
and the constraints (\ref{12}).

The Lagrange function is equal to
\be
L=(\dot{x} p) + (\dot{n} p^{(n)})=(\dot{x} \pi) + (\dot{n} p^{(n)})
+e (A\dot{x})\ ,
\label{27}
\ee
where the momenta $\pi_\mu$ and $p^{(n)}_\mu$ should be expressed in terms
of the generalized coordinates and velocities using Eqs.(\ref{26}). Since
\begin{eqnarray*}
(\dot{x} \pi)=2\Lambda \pi^2=2\Lambda 
(e F_{\mu\nu} S^{\mu\nu} + m^2)\ ,\\
(\dot{n} p^{(n)})=2\Lambda e F_{\mu\nu}n^\nu p^{(n)\mu}=-\Lambda 
e F_{\mu\nu} S^{\mu\nu}\ ,
\end{eqnarray*}
then we have
\\
\be
L=2m^2\Lambda \left( 1+\frac{e}{2m^2} F_{\mu\nu} S^{\mu\nu}\right) +
e (A\dot{x})\ .
\label{28}
\ee
\\

Up to terms linear in the electromagnetic field strength we find from 
Eqs.(\ref{26}) and (\ref{12}):
\be
\Lambda = \frac{(\dot{n}\dot{x})}{2m\sqrt{\dot{n}^2}}+ O(e^2F^2)\ ,
\label{29}
\ee
and thus we obtain finally
\\
\be
L=\frac{m (\dot{n}\dot{x})}{\sqrt{\dot{n}^2}}\left( 1+\frac{e}{2m^2}
F_{\mu\nu} S^{\mu\nu}\right) + e(A \dot{x}) + O(e^2F^2)\ ,
\label{30}
\ee
where
\\
\be
S^{\mu\nu}=\frac{m}{\sqrt{\dot{n}^2}}\left( n^\mu \dot{x}^\nu-
n^\nu \dot{x}^\mu\right) -\frac{m (\dot{n}\dot{x})}{(\dot{n}^2)^{3/2}}
\left(  n^\mu \dot{n}^\nu- n^\nu \dot{n}^\mu\right)\ .
\label{31}
\ee
\\

It is worth mentioning that in the absence of an external electromagnetic
field from the equations of motion (\ref{20}) it follows that 
$\dot{n}^\mu=0$. In this case it is more convenient to adopt a first
order formalism (\eg as advocated in \cite{faddeev,jackiw}) and 
take the Lagrangian as 
$$
L=(\dot{x} p) + (\dot{n} p^{(n)}) - \Lambda (p^2-m^2) - \lm_1 (p p^{(n)})
- \lm_2 (p n)\ .
$$

We have presented the Hamiltonian description for the motion 
of a particle with gyromagnetic ratio $g=2$ in an external electromagnetic 
field. The main ingredients of the model are the following: the phase
space of the system is given by the coordinates $x^\mu$ and $n^\mu$ and
their conjugated momenta $p_\mu$ and $p^{(n)}_\mu$ respectively;
the antisymmetric spin matrix $S_{\mu\nu}$ is defined in terms of the 
four vectors $n_\mu$ and $p^{(n)}_\mu$ (see Eq.(\ref{2})) by analogy with 
the orbital angular momentum tensor; the total Hamiltonian is given in
(\ref{14}) with the primary constraints defined in (\ref{12}). Out of
these starting points we have obtained the equations of motion (\ref{21})
and (\ref{22}) (or equivalently Eq. (\ref{25})).

Let us now consider the case of a particle with gyromagnetic ratio 
$g \neq 2$. A straightforward generalization of our previous ($g=2$)
analysis to the case $g\neq 2$ would be the introduction of the
factor $g/2$ in the 
 interaction term $e F_{\mu\nu} S^{\mu\nu}$ of the constraint $\Phi$ 
in Eqs.(\ref{12}). However, in order to
keep the latter constraint as first-class and the constraints 
$\varphi_1$ and
$\varphi_2$ as second-class, we should further modify the constraint
$\Phi$.  Taking the linear combination
$$
\tilde{\Phi }= \pi^2-\frac{eg}{2} F_{\mu\nu} S^{\mu\nu} - m^2 +
A \varphi_1 + B \varphi_2\ ,
$$
and imposing the conditions $\{\tilde{\Phi} ,\varphi_1\}=
\{\tilde{\Phi} ,\varphi_2\}=0 $ for the constraint $\tilde{\Phi}$ to be
first-class, we can determine the coefficients $A$ and $B$. Finally we
find
\\
\be
\tilde{\Phi }= \pi^2-\frac{eg}{2} F_{\mu\nu} S^{\mu\nu} - m^2 -
\frac{e(g-2)}{(\pi^2+ \frac{e}{2} F_{\mu\nu} S^{\mu\nu})} 
F^{\alpha \beta } \pi _\alpha  S_{\beta \gamma }\pi^\gamma\ .
\label{32}
\ee
\\
 
The equations of motion which follow from the new Hamiltonian
\be
H= \Lambda \tilde{\Phi} +\lm_1\varphi_1+\lm_2\varphi_2\ 
\label{33}
\ee
(with $\varphi_1$ and $ \varphi_2$ defined in (\ref{12})) read as
follows:
\\
\bea
\dot{x}^\mu=2\Lambda \pi^\mu\ - 
\frac{\Lambda e(g-2)}{(\pi^2+ \frac{e}{2} F_{\mu\nu} S^{\mu\nu})} 
F_{\alpha \beta } \pi ^\alpha  S^{\beta \mu } \ ,\CR
\dot{n}^\mu=\Lambda  eg F^{\mu\nu} n_\nu\ - 
\frac{\Lambda e(g-2)}{(\pi^2+ \frac{e}{2} F_{\mu\nu} S^{\mu\nu})} 
F^{\alpha \beta } \pi _\alpha n_\beta \pi^\mu  \ ,\CR
\dot{\pi}_\mu=2\Lambda e F_{\mu\nu} \pi^\nu\ - 
\frac{\Lambda e^2(g-2)}{(\pi^2+ \frac{e}{2} F_{\mu\nu} S^{\mu\nu})} 
F_{\alpha \beta } \pi ^\alpha F_{\mu\nu} S^{\beta \nu } \ ,\CR
\dot{p}^{(n)}_\mu=\Lambda eg F_{\mu\nu} p^{(n) \nu} -
\frac{\Lambda e(g-2)}{(\pi^2+ \frac{e}{2} F_{\mu\nu} S^{\mu\nu})} 
F^{\alpha \beta } \pi _\alpha p^{(n)}_\beta \pi_\mu \ .
\label{34}
\eea
\\

For the spin tensor $S_{\mu\nu}$ and vector $S_\mu$ we get in turn:
\bea
\dot{S}_{\mu\nu}=&-&\Lambda eg F^{\alpha \beta }\left( 
S_{\mu\alpha}
g_{\nu \beta }+S_{\nu \beta } g_{\mu \alpha }\right)  \CR
            &+& \frac{\Lambda e(g-2)}{(\pi^2+ \frac{e}{2} F_{\mu\nu} 
S^{\mu\nu})} \pi_\alpha F^{\alpha \beta }\left( S_{\beta\mu}\pi_\nu
+ S_{\nu\beta}\pi_\mu \right) \ ,\CR
\dot{S}_\mu=&&\Lambda eg F_{\mu \nu }S^{\nu}
- \frac{\Lambda e(g-2)}{\pi^2}\pi_\mu F_{\alpha \beta }\pi^\alpha 
S^{\beta} \CR
&-&\frac{\Lambda e^2 (g-2)}{(\pi^2)^{3/2} (\pi^2+ \frac{e}{2} F_{\mu\nu} 
S^{\mu\nu})}\pi_\mu \varepsilon_{\alpha\beta\gamma\delta} F^{\alpha
\alpha'} S_{\alpha'} F^{\beta\beta'} \pi_{\beta'} S^\gamma \pi^\delta \ .
\label{35}
\eea
  
We note that from the equations of motion (\ref{34})-(\ref{35}) it 
follows that the value of the spin (\ref{19}) is preserved in time. 
Moreover, using the expression for $\pi^\mu$ which follows from the first
equation in (\ref{34}), \ie
$$
\pi^\mu=\frac{1}{2\Lambda}\left( \dot{x}^\mu+ 
\frac{\Lambda e(g-2)}{(\pi^2+ \frac{e}{2} F_{\mu\nu} S^{\mu\nu})} 
F_{\alpha \beta } \pi ^\alpha  S^{\beta \mu }\right) \ ,
$$
and substituting it into the definition of the spin vector (\ref{17}),
it is easy to show that
\be
S_\mu=\frac{1}{4\Lambda\sqrt{\pi^2}}\varepsilon_{\mu \alpha \beta \gamma }
\dot{x}^\alpha S^{\beta \gamma }\ .
\label{36}
\ee 
Thus the condition $S_\mu \dot{x}^\mu=0$ is automatically satisfied
by the equations of motion.   

For the case $g=2$, if one chooses 
$\Lambda=1/2m$, Eqs.(\ref{34})-(\ref{35}) coincide with Eqs.(\ref{21}) 
and (\ref{25}). However, when $g \neq 2$, additional quadratic and 
higher order terms in the field strength $F_{\mu\nu}$ appear. These terms 
reproduce the Bargmann-Michel-Telegdi equation \cite{BMT} in the linear
approximation. Indeed, from (\ref{34})-(\ref{35}) we obtain
\bea
\ddot{x}_\mu=\frac{e}{m}F_{\mu\nu} \dot{x}^\nu + (g-2)O(e^2F^2)\ ,\CR
\dot{S}_{\mu}=\frac{eg}{2m} F_{\mu \nu } S^{\nu} 
-\frac{e(g-2)}{2m}\dot{x}_\mu F_{\alpha\beta}\dot{x}^\alpha S^\beta
+ (g-2)O(e^2F^2)\ .
\label{37}
\eea

We remark that the extra  terms of order $e^2 F^2$ in Eqs.(\ref{37}) 
do not coincide with the ones previously obtained in the literature  
\cite{anandan, bander, costella}. One way to see that is to consider the
nonrelativistic limit. For particles with $g=2$ our equations do not
predict any deviation from the standard Lorentz force, while in Refs.
\cite{anandan, bander, costella} a term proportional to $g \vec{B}\times
(\vec{S}\times \vec{E})$ is obtained. For $g \neq 2$, however, 
Eqs.(\ref{34}) do predict in the nonrelativistic limit an additional  
force, namely,  
$$
-\frac{e^2(g-2)}{2m^3} \vec{B}\times(\vec{S}\times \vec{E}) \ ,
$$ 
which modifies the Lorentz force.  

We have considered so far the case when the electromagnetic field
strength $F_{\mu\nu}$ is constant. We 
can modify the model to obtain the equations of motion in slowly 
varying fields. In the latter case, the constraint $\Phi$ defined in 
Eq.(\ref{9}) as well as $\tilde{\Phi}$ in Eq.(\ref{32}) will also 
generate the correct equations of motion so that in the nonrelativistic 
limit we obtain the term $\frac{eg}{2m} \vec{\nabla} (\vec{B}\cdot 
\vec{S})$ \cite{anandan, bander, costella}.

\vskip 1.5cm

We are grateful to M. Bander, I. Khriplovich and G. Kunstatter for 
useful discussions.

\end{document}